# Measurement of Heavy Flavor and Quarkonia Production by PHENIX

M. L. Brooks for the PHENIX Collaboration
*LANL, Los Alamos, NM 87545, USA*

Heavy quarks (charm and bottom) are good probes of the hot and dense medium created in relativistic heavy ion collisions since they are mainly generated at the beginning of collisions and interact with the media in all collision stages. In addition, heavy flavor quarkonia production is thought to be uniquely sensitive to the deconfined medium of the Quark Gluon Plasma (QGP) through color screening. Heavy quark production has been studied by the PHENIX experiment at RHIC via measurements of single leptons from semi-leptonic decays in both the electron channel at mid-rapidity and in the muon channel at forward-rapidity. Large suppression and azimuthal anisotropy of single electrons have been observed in Au+Au collisions at $\sqrt{s}$ = 200 GeV. These results suggest a large energy loss and flow of the heavy quarks in the hot, dense matter. The PHENIX experiment has also measured J/ψ production up to $\sqrt{s}$ = 200 GeV in p+p, d+Au, Cu+Cu and Au+Au collisions, both at mid ($|y|<0.35$) and forward ($1.2<|y|<2.2$) rapidities. In the most energetic collisions (central Au+Au), more suppression is observed at forward rapidity than at central rapidity. This can be interpreted either as a sign of quark recombination, or as a hint of additional cold nuclear matter effects. The current status and understanding of all these measurements will be reviewed.

Because charm and beauty production takes place in the early, hard interactions in A+A collisions, these heavy quarks are available to probe the medium throughout the collision process. Heavy flavor measurements can allow us to understand the quark-mass dependence of energy loss in the medium, and to understand whether color screening can significantly modify quarkonia production in A+A collisions. However, there are a number of possible in-medium effects which can modify the production in A+A collisions compared to p+p collisions, and unraveling these requires a systematic study of open and closed heavy flavor production, production versus different system sizes and centrality, and mapping out the differential production versus rapidity and $p_T$.

The probability for a hard interaction to occur can be modified even before in-medium effects take place if the parton distribution functions are effectively modified in a nucleus. Because much of the heavy flavor production at RHIC comes from gluon-gluon fusion (especially for quarkonia production), the heavy flavor production is especially sensitive to any modification to the gluon distribution function such as would come from shadowing [i]. Since shadowing is a cold nuclear matter effect, one can attempt to separate this effect from hot nuclear matter effects by measuring particle production in d+Au collisions and comparing to p+p collisions.

Quarkonia production can also be *enhanced* in A+A collisions if there is enough heavy flavor production in each A+A collision to allow for the possibility of recombination of a c and cbar pair from two independent hard collisions into a quarkonia meson. If recombination is a significant contributor to quarkonia production, you would expect the enhancement in production to increase versus the number of binary collisions, and the $p_T$ and rapidity distributions should be narrowed to reflect the convolution of two independent distributions [ii].

As the quarks propagate through the medium, they will suffer energy loss. However, radiative and collisional energy loss as well as collisional dissociation can all potentially contribute to suppression in hot nuclear matter, and the partons can suffer energy loss in cold nuclear matter. Since the various energy loss mechanisms typically result in different kinematic dependencies, one can help separate the mechanisms by mapping out the particle production modification versus quark mass, rapidity, and $p_T$, and measuring production separately in d+A and A+A collisions.

Particle production of both open heavy flavor and quarkonia can also be modified by break-up of the mesons inside the hot dense medium or by color screening. J/ψ mesons appear to suffer from a measurable break-up cross section even in cold nuclear matter [iii], and this can be deduced by comparing particle production in d+A collisions with that in p+p collisions. In a quark gluon plasma, the c-cbar are expected to be screened if the plasma reaches high enough temperature, and the following J/ψ formation suppressed. The amount that screening affects production in A+A collisions can be deduced by measuring both open and heavy flavor production (open heavy flavor production will not suffer screening effects but will share many of the other in-medium modification effects with quarkonia), by measuring production in d+A and A+A collisions to pull out cold nuclear matter effects, and by measuring the production of more than one quarkonia, e.g. J/ψ, ψ', Y, $\chi_c$, each will see the onset of screening at different temperatures. It has also been shown that because of their short formation time, D and B mesons will have a significant probability to break-up inside the quark gluon plasma, undergo additional interactions, and then reform before escaping the medium [iv]. This interaction will result in a $p_T$ spectrum which is shifted to lower energies, just as radiative energy loss causes a $p_T$ shift in produced spectra. However, because of the different formation times of D and B mesons, this type of effective energy loss can be separated from radiative energy loss contributions by measuring separately D and B production versus $p_T$.

Open heavy flavor production is measured in PHENIX primarily through the measurement of inclusive electrons or muons. For the electron measurement, the electrons that come from meson decay or photon conversions are measured and subtracted from the inclusive spectrum and the remainder is attributed to electrons coming from the semi-leptonic decay of D and B mesons [v]. This remainder is also referred to as the non-photonic electron component. For the inclusive muon measurement, the muons that come from meson decay are inferred from the vertex distribution of the inclusive spectrum, the hadronic punch-through component is inferred by measuring the spectra versus various absorber lengths, and the two components are subtracted from the inclusive single particle spectra. As in the electron case, the remainder is attributed to muons coming from the semi-leptonic decay of D and B mesons [vi].

Quarkonia measurements are made in PHENIX either by detecting opposite sign electrons at central rapidity (|y|<0.35) or by detecting opposite sign muons at forward rapidity (1.2<|y|<2.4), reconstructing the invariant mass of the di-lepton pair, and subtracting the continuum background [vii].

PHENIX has measured the modification of open heavy flavor production in Au+Au collisions by measuring the non-photonic single electron spectrum in Au+Au collisions and comparing this to the production in p+p collisions, scaled up by the number of binary collisions calculated for a given centrality. This $R_{AA}$ ratio is shown in Figure 1 versus $p_T$ for the 0-10% most central collisions. As can be seen, there is a significant suppression of particle production at high to moderate $p_T$ compared to the production expected from p+p measurements. The suppression measured is comparable to the suppression measured for light quark mesons even though models of radiative energy loss, which reproduced the light meson spectra very well, predicted little energy loss for heavy quarks. Significant flow was also measured for the inclusive single electron spectrum, as seen in Figure 1. These results imply that heavy flavor is suffering from significant in-medium interactions which cause a shifting of the $p_T$ spectra as well as flow. Several models now attempt to reproduce this data [iv,viii], but with different mechanisms (radiative energy loss, collisional energy loss, collisional dissociation) accounting for the suppression. Additional measurements which separate the c and b contributions to the spectra could allow us to determine which mechanisms are causing the large suppression.

The charm and beauty components have been separated in PHENIX by fitting the invariant mass spectrum of e-hadron correlated tracks [ix]. The spectral functions are different for e-h coming from D decays than for those coming from B-decays, allowing a separation of the two. The resulting b/c ratio that is extracted from p+p data is shown in

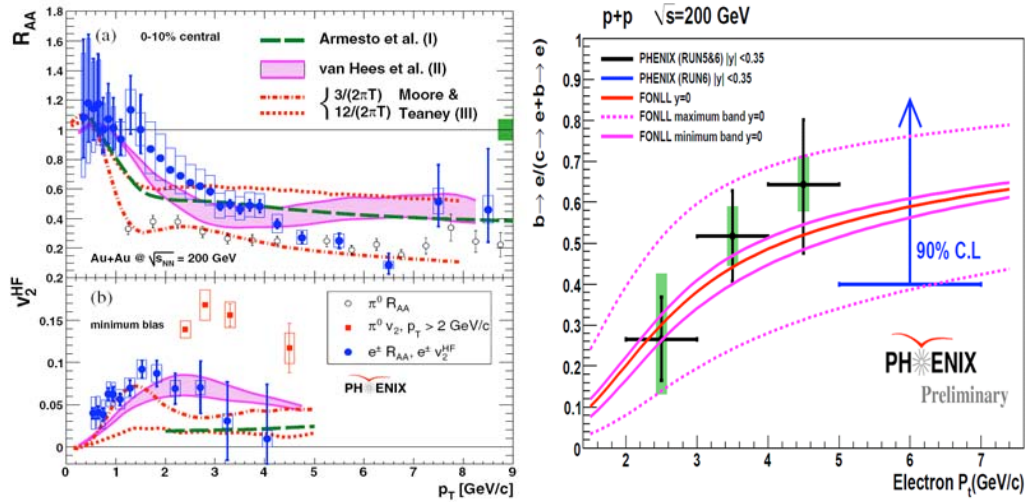

**Figure 1** The nuclear suppression factor, $R_{AA}$, for 0-10% most central Au+Au collisions, measured by PHENIX for non-photonic electrons, versus $p_T$ (upper left). The $v_2$ for non-photonic electrons, in Au+Au collisions is also shown (lower left). The extracted b/c ratio for the single electron spectrum, versus $p_T$ is shown on the right.

Figure 1 B-decays, allowing a separation of the two. The resulting b/c ratio that is extracted from p+p data is shown in Figure 1 versus $p_T$. As can be seen, the beauty contribution appears to begin to dominate at approximately 3-4 GeV, albeit with large error bars. With the PHENIX upgrade detectors, we expect to significantly improve the ability to separate these components via detached verrtices and clearly distinguish energy loss mechanisms contributing to heavy flavor suppression.

PHENIX has also measured J/ψ production in p+p, d+Au, Cu+Cu and Au+Au collisions. We present in Figure 2 the nuclear modification factor, $R_{AA}$, measured for J/ψ production in Au+Au collisions at √s = 200 GeV at central rapidity (left panel) and forward rapidity (right panel). The theoretical prediction for the suppression expected just from cold nuclear matter effects (based on d+Au measurements), is also shown in the panels with yellow bands. As can be seen, the measured spectrum at central rapidity is consistent with cold nuclear matter effects but the forward rapidity suppression is beyond the 1-σ expectation for cold-nuclear matter suppression alone. An increased suppression at forward rapidity is not expected from hot nuclear matter effects (they should decrease as you move to forward rapidity and go through less medium), so it is postulated that either there are additional cold-nuclear matter effects adding to the suppression at forward rapidity or perhaps recombination is making a significant contribution to J/ψ production. Recombination would produce more J/ψs at central rapidity than at forward rapidity, therefore potentially raising the suppression factor at central rapidity above the forward rapidity $R_{AA}$. To better understand the suppression mechanisms at play, we plan to improve the cold nuclear matter predictions by analyzing Run-8 d+Au data, improve the open heavy flavor measurements so that the recombination contributions can be better constrained, and add additional vector meson measurements to the J/ψ measurement so we can map out the suppression versus binding energy.

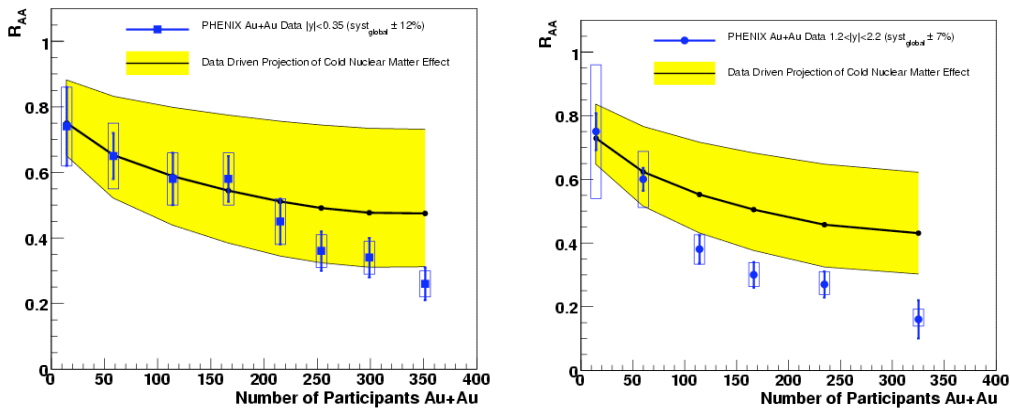

**Figure 2** $R_{AA}$ in Au+Au collisions for J/ψ at central rapidity (left) and forward rapidity (right).

With the PHENIX silicon vertex tracking upgrade detectors, we expect to significantly improve the open heavy flavor and vector meson measurement capabilities in PHENIX. With the addition of precision vertex measurements, we will be able to much more cleanly separate the leptons from D and B decay from backgrounds. With this precise separation, we will be able to significantly reduce the systematic error bars on our single lepton measurements. We can also separately tag charm and beauty decay leptons because of the different life-times of D and B mesons. Taken together, we will be able to make $R_{AA}$ measurements such as those shown in Figure 3, left-hand panel. The vertex tracking will also bring improved mass resolutions in both our di-electron and di-muon channels, and allow for a clean separation of J/ψ from ψ'. Our expected vector meson performance, with one year of running at RHIC II luminosities, is shown in the right-hand panel of Figure 3.

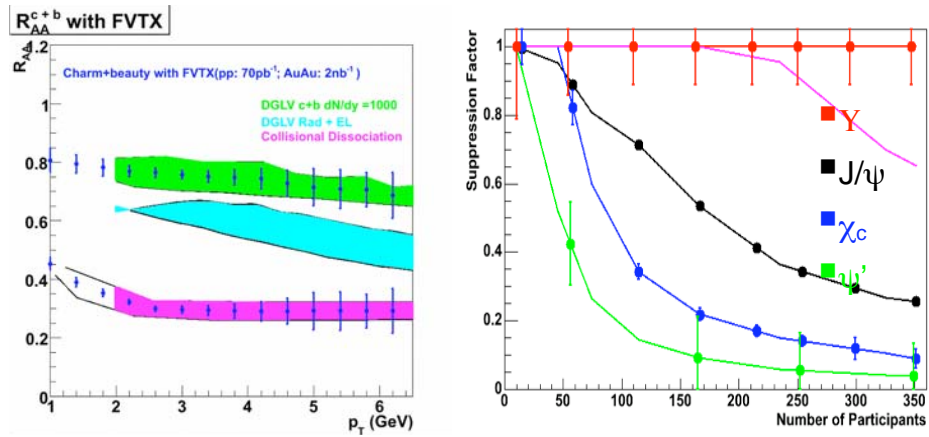

**Figure 3** Charm and beauty $R_{AA}$ measured with upgrade silicon detectors (left) for one year of RHIC I luminosity (left) and vector meson measurement capabilities in 1 RHIC II year.

**Acknowledgements**

Work supported by the U.S. Department of Energy.